\documentclass[aps,longbibliography,notitlepage,floatfix,11pt,tightenlines,nofootinbib]{revtex4-1}

\usepackage{bm,amsmath,amssymb,graphicx,stmaryrd,subfigure}
\usepackage[abs]{overpic}
\newcommand{\uvc}[1]{\bm{\mathrm{\hat #1}}} 

\usepackage[usenames,dvipsnames]{xcolor}

\newcommand{\bX}{{\bf X}}

\begin{document}

\title{A conserved quantity in thin body dynamics}
\author{J A Hanna}
\email{hannaj@vt.edu}
\affiliation{
Department of Biomedical Engineering and Mechanics,\\
Department of Physics,\\
Virginia Polytechnic Institute and State University, Blacksburg, VA 24061, U.S.A.}
\author{H Pendar}
\affiliation{Department of Biomedical Engineering and Mechanics,\\
Virginia Polytechnic Institute and State University, Blacksburg, VA 24061, U.S.A.\\}

\date{\today}

\begin{abstract}
Thin, solid bodies with 
metric symmetries admit a restricted form of reparameterization invariance. Their dynamical equilibria include 
motions with both rigid and flowing aspects.  On such configurations, a quantity is conserved along the intrinsic coordinate corresponding to the symmetry. 
 As an example of its utility, this conserved quantity is combined with linear and angular momentum currents to construct solutions 
for the equilibria of a rotating, flowing string, for which it is akin to Bernoulli's constant.
\end{abstract}

\maketitle



The mechanics of thin bodies are amenable to a description in which energies are effectively expressed with an integral over a surface or curve.  On sufficiently long time scales, and when longitudinal 
waves 
are not important, the effective action is often taken to be metrically constrained, with a low-dimensional body stress tensor enforcing local distances between moving pieces of material.  This approach, either at the level of the action or equations of motion, has been used to describe systems at many scales, from the overdamped motions of flagella \cite{GoldsteinLanger95, Lauga07}, filaments \cite{Hinch76, Powers10}, or macromolecules \cite{EdwardsGoodyear72}, to the inertial dynamics of elastic rods \cite{WhitmanDeSilva69, ColemanDill92, GorielyTabor96, LinRavi-Chandar06, Bergou08}, soft robot links \cite{DavisHirschorn94}, yarn \cite{BatraFraser15}, flags \cite{ShelleyZhang11}, or sonar and telegraph cables in the ocean \cite{Zajac57, Dowling88-1, Goyal05}. 

\begin{figure}[h]
	\begin{overpic}[width=3in]{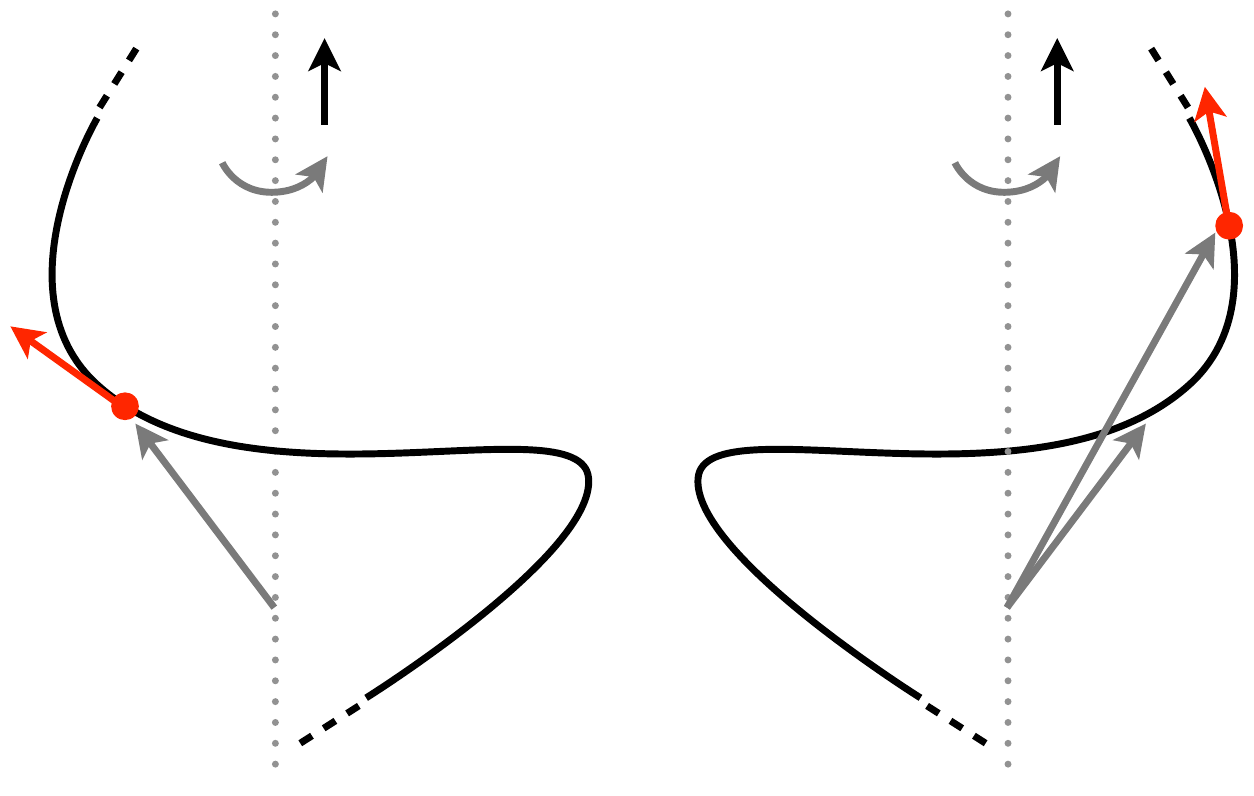}
	\put(30,100){$\omega$}
	\put(60,120){$\uvc{z}$}
	\put(157,100){$\omega$}
	\put(187,120){$\uvc{z}$}
	\put(176,90){$\bX(s, t)$}
	\put(195,50){$\small{\bX(\eta,t)}$}
	\put(213,112){$T\partial_s\bX$}
	\end{overpic}
\caption{Two snapshots of a one-dimensional body $\bX$ rotating with angular velocity $\omega\uvc{z}$ and carrying a uniform tangential flow of material described by the velocity field $T\partial_s\bX$, $T$ constant.  Shown are the material description $\bX(s,t)$, which follows the flow on the body, and the body-Eulerian description $\bX(\eta,t)$, $\eta \equiv s + Tt$, which rotates rigidly.  The two descriptions are identical if $T=0$.  Let the straight grey arrow on the left snapshot denote either description at some point in time $t_1$.  The snapshot on the right occurs at a later time $t_2$; the material position vector has moved along the curve, while both curve and body-Eulerian position vector have rigidly rotated.  At this later time, the $s$ value corresponding to the same $\eta$ value is smaller by an amount $T(t_2-t_1)$; it corresponds to a piece of material that came from a location farther back on the curve.  The snapshots correspond to the same range of $\eta$ values, but different ranges of $s$ values.}
\label{rot}
\end{figure}
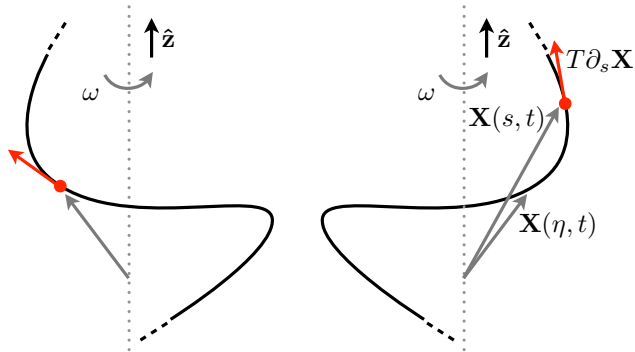

Thin bodies such as rods and solid membranes moving in a higher-dimensional space share characteristics with both rigid bodies and fluids.  The metric constraints of inextensible solids preclude the broad reparameterization invariance inherent to fluids and fluid membranes, leaving only highly restricted reparameterization invariance associated with any metric symmetries the body may possess.  Effectively one has global, rather than local, symmetries on the body.  However,  these intrinsic symmetries are not trivially identifiable with rigid transformations of embedding coordinates, because a lower-dimensional body can adopt extrinsic curvature in the surrounding space.  Thus, unlike space-filling rigid continua, the body may display additional material flows in equilibrium in addition to, and distinct from, rigid motions.  Simple examples include a string being unspooled \cite{BatraFraser15}, a cable being laid down on the ocean floor \cite{Zajac57}, 
or the flowing skirt of a rotating dancer \cite{Guven13skirts}.  Figure \ref{rot} is a schematic of such a combined rigid-flowing configuration.

We will describe such a system using several types of coordinates.  Aside from a time coordinate, there are three important sets of spatial coordinates to consider.
First, there are time-independent coordinates attached to pieces of material.  For an inextensible body, these can also be associated with geometric parameters, such as arc length.
Second, there are coordinates for which material motions tangent to the body have been removed.  These are effectively Eulerian from the intrinsic viewpoint of the body, but they are attached to a body moving through an embedding space, so have some Lagrangian character as well.  This viewpoint is called the ``orthogonal gauge'' in \cite{Brower84}.  Finally, there are the embedding coordinates, which we will assume to be the simple Cartesian coordinates associated with an inertial frame in $\mathbb{E}^3$.
We will refer to these three sets of spatial coordinates as material or body (Lagrangian), body-Eulerian, and embedding (Eulerian), respectively.

The action of a continuum may possess symmetries in any of these coordinates, and corresponding conserved quantities may be extracted by application of Noether's theorems.  
Translational and rotational symmetries with respect to embedding coordinates correspond to familiar conservation laws for linear and angular momentum.  For solids, symmetries with respect to material coordinates imply symmetries of the body metric. 
For space-filling inextensible solids \cite{Eshelby75, KienzlerHerrmann00, Maugin11}, these must necessarily coincide with rigid motions of the embedding coordinates.  Fluids have a greater range of allowable reparameterizations or ``relabelings'', namely any transformations that conserve the density of labels \cite{Newcomb67, Bretherton70, Salmon88, PadhyeMorrison96-1}. 
Some workers have investigated symmetries of thin solid-like or fluid-like bodies in a higher-dimensional background, including relativistic extended objects \cite{Carter94, Carter01, Arreaga00, Armas13}, energy-bearing curves \cite{Capovilla02}, or fluid membranes \cite{CapovillaGuven02}.

In this note, we introduce a conserved quantity for thin body dynamical equilibria, corresponding to a symmetry with respect to either body or body-Eulerian coordinates. 
In the former case, which corresponds to rigid motions, the quantity is identical to a flux derived in a non-variational manner for uniform rods by Maddocks and Dichmann \cite{MaddocksDichmann94}.  In either case, it is a useful tool in the search for solutions.
 We illustrate these ideas with the concrete example of a rotating, flowing string, known as a ``yarn balloon'' in textile processing applications \cite{BatraFraser15}.   In this example, the conserved quantity is constructed from portions of the kinetic energy and body stress.  Related problems involving axially moving structures such as belts have stimulated discussion of Eulerian action principles and the boundary conditions they produce \cite{Renshaw98}.  
Other related work seeks to extend action principles and Noether's theorems to general non-material volumes \cite{McIver73, IrschikHoll02, CasettaPesce13, Casetta15}.

 Without loss of generality, we will discuss a one-dimensional body such as that of Figure \ref{rot}.  
Let us begin by describing the body with a Cartesian embedding vector $\bX(s,t)$, parameterized by a material coordinate $s$ and time $t$.  Consider the action of a classical string, which is an integral over a constrained kinetic energy density
\begin{equation}
\mathcal{L} = \tfrac{1}{2}\mu\partial_t\bX\cdot\partial_t\bX - \tfrac{1}{2}\sigma\left(\partial_s\bX\cdot\partial_s\bX - 1\right) \, .
\end{equation} 
Here, $\mu$ is a uniform mass density, and the stress or tension $\sigma$ is a multiplier enforcing the metric constraint $\partial_s\bX\cdot\partial_s\bX = 1$.  The string is inextensible, and the material coordinate is also the arc length.  The rather trivial metric of the body, $\partial_s\bX\cdot\partial_s\bX$, is symmetric in $s$.

The variation of body coordinates will be performed in an analogous manner to the time variation that leads to the temporal boundary term representing the total energy (Hamiltonian) of a classical action.  Recall that for an action of the form
\begin{equation}
	A = \int_{t_1}^{t_2} dt \,L = \int_{t_1}^{t_2} \!\!dt \int_{s_1}^{s_2} \!\!ds \,\mathcal{L}\left(\bX(s,t)\right) \, ,
\end{equation}
a time shift $t \rightarrow t + \delta t$ results in a new action
\begin{align}
	A + \delta A &=  \int_{t_1}^{t_2} \!\!d(t+\delta t) \int_{s_1}^{s_2} \!\!ds \,\mathcal{L}\left(\bX(s,t+\delta t)\right)  \, , \\
	&=  \int_{t_1 - \delta t}^{t_2 - \delta t} \!\!dt \int_{s_1}^{s_2} \!\!ds \,\mathcal{L}\left(\bX(s,t+\delta t)\right)  \, , \\
	&= A \left.-L\right|_{t=t_1}^{t=t_2}\delta t +  \int_{t_1}^{t_2} \!\!dt \int_{s_1}^{s_2} \!\!ds \,\frac{\delta\mathcal{L}}{\delta\bX}\cdot \partial_t\bX(s,t)\delta t  + \mathcal{O}\left(\delta t^2\right)\, .
\end{align}
After integration by parts, the integral term may contribute to the temporal boundary term.  In our case,
\begin{align}
	\delta A &= \left.-\int_{s_1}^{s_2} \!\!ds \,\tfrac{1}{2}\mu\partial_t\bX\cdot\partial_t\bX\right|_{t=t_1}^{t=t_2}\delta t +  \int_{t_1}^{t_2} \!\!dt \int_{s_1}^{s_2} \!\!ds \, \mu\partial_t\bX\cdot\partial_t\left(\partial_t\bX \delta t\right) - \sigma\partial_s\bX\cdot\partial_s\left(\partial_t\bX \delta t\right) \, , \\
	&= \left.-\int_{s_1}^{s_2} \!\!ds \,\tfrac{1}{2}\mu\partial_t\bX\cdot\partial_t\bX\right|_{t=t_1}^{t=t_2}\delta t +  \int_{t_1}^{t_2} \!\!dt \, \partial_t \left( \int_{s_1}^{s_2} \!\!ds \, \mu\partial_t\bX\cdot\partial_t\bX \delta t \right) + \mathrm{other\, terms} \, .
\end{align}
We are only interested in the temporal boundary term, which has contributions from the shift in endpoints as well as an integration by parts of the kinetic term inside the integral.\footnote{The variation also yields bulk and spatial boundary terms that, for the present purely mechanical system, provide redundant information with respect to that obtained from variation of the embedding vector $\bX$.}  This boundary term can be set equal to a constant,
\begin{equation}
	E = \int_{s_1}^{s_2} \!\!ds \,\tfrac{1}{2}\mu\partial_t\bX\cdot\partial_t\bX \, .
\end{equation}
This is the total energy of the body, purely kinetic in this case because the constraints store no energy.  

As $E$ is an integral over the entire body, it is not very useful in solving for the local dynamics of an extended object like a string.  
Instead, we would like to find quantities that are uniform along the string, that is, conserved along a body coordinate rather than in time.  Now consider an analogous variation of the material coordinate $s \rightarrow s + \delta s$.  We find
\begin{align}
	\delta A &= \int_{t_1}^{t_2} \!\!dt \left[ \left. -\mathcal{L}\right|_{s=s_1}^{s=s_2}\delta s +  \int_{s_1}^{s_2} \!\!ds \,\frac{\delta\mathcal{L}}{\delta\bX}\cdot \partial_s\bX(s,t)\delta s  \right] \, , \\
	&= -\int_{t_1}^{t_2} \!\!dt \left[ \left. \tfrac{1}{2}\mu\partial_t\bX\cdot\partial_t\bX\right|_{s=s_1}^{s=s_2}\delta s + \int_{s_1}^{s_2} \!\!ds \, \partial_s\left( \sigma\partial_s\bX\cdot\partial_s\bX \delta s \right) \right] + \mathrm{other\, terms} \, .
\end{align}
If the system is symmetric in a spatial coordinate, the corresponding spatial boundary term will be a function of time that is uniform along the string.  In the current context, that of dynamical equilibria, these functions of time will be proper constants, and will be referred to as such in what follows. For rigid body motion of $\bX(s,t)$, a uniform reparameterization $\delta s$ is a symmetry of the action, and yields the constant
\begin{equation}
	C_s = \tfrac{1}{2}\mu\partial_t\bX\cdot\partial_t\bX + \sigma \label{Cs} \, .
\end{equation}
This is identical to the constant $c_3$ in \cite{Hanna13}, and to the constant $\tfrac{1}{2}C$ in \cite{Guven13skirts} if $\sigma$ is replaced with the appropriate component of the surface stress there.  
It is a particular manifestation of a conservation law for uniform, hyperelastic rods \cite{MaddocksDichmann94}.

We may also vary the position vector, as if to derive momentum balance equations.  After $\bX \rightarrow \bX + \delta\bX$, we have
\begin{align}
	\delta A &=  \int_{t_1}^{t_2} \!\!dt \int_{s_1}^{s_2} \!\!ds \,\frac{\delta\mathcal{L}}{\delta\bX}\cdot \delta \bX \, , \\
	&= \int_{t_1}^{t_2} \!\!dt  \int_{s_1}^{s_2} \!\!ds \, \partial_s\left( \sigma\partial_s\bX\cdot\delta\bX \right)+ \mathrm{other\, terms} \, .
\end{align}
If $\delta \bX$ corresponds to a continuous symmetry of the system, we may obtain new constants.  For a system rigidly rotating about an axis $\uvc{z}$, there are two such symmetries, namely translations $\uvc{z}\delta z$ along $\uvc{z}$ and infinitesimal rotations $(\uvc{z} \times \bX)\delta\theta$ about $\uvc{z}$. Thus,
\begin{align}
	P_z &= \sigma\partial_s\bX\cdot\uvc{z} \, , \label{Pz} \\
	J_z &= \sigma\partial_s\bX\cdot\left(\uvc{z} \times \bX\right) \, . \label{Jz}
\end{align}
These constants were called $c_1$ and $c_2/\omega$ in \cite{Hanna13}, and $J_z$ here is analogous to $\tfrac{1}{2}J_z$ in \cite{Guven13skirts}.

We may now easily obtain a quadrature for rigidly rotating string equilibria.  Note that we can form an orthonormal triad of $\uvc{z}$, $\uvc{\theta} \equiv \frac{\uvc{z} \times \bX}{\|\uvc{z} \times \bX\|}$, and $\uvc{r} \equiv -\uvc{z} \times \uvc{\theta}$, and that with the origin of $\bX$ on the $\uvc{z}$ axis, $\bX \equiv r\uvc{r} + z\uvc{z}$ and, thus, $\uvc{z} \times \bX = r\uvc{\theta}$ and $\partial_s\bX\cdot\uvc{r} = \partial_s r$.  For a string rigidly rotating with angular frequency $\omega$ about $\uvc{z}$, $\partial_t\bX = \omega r \uvc{\theta}$, so $\sigma$ may be written as a function of a single variable $r = \|\uvc{z} \times \bX\|$.  Now use the three constants \eqref{Cs}, \eqref{Pz}, and \eqref{Jz} to write:
\begin{align}
	(\partial_s\bX\cdot\uvc{r})^2 &= 1 - (\partial_s\bX\cdot\uvc{z})^2 - (\partial_s\bX\cdot\uvc{\theta})^2 \, , \label{sumsq} \\
	\left(\partial_s r\right)^2 &= 1 - \frac{P_z^2 + (J_z/r)^2}{\left(C_s - \tfrac{1}{2}\mu\omega^2r^2\right)^2} \, . \label{quadrigid}
\end{align}
This can also be reformulated in terms of an elliptic integral in $r^2$ \cite{Hanna13}, but we need not do that here.  Solving for $r$, one then obtains $\sigma$ directly from \eqref{Cs}, and the other two coordinates by quadratures using \eqref{Pz} and \eqref{Jz} or \eqref{sumsq}.  Nondimensionalizing \eqref{quadrigid}, using some chosen length scale along with $\mu$ and $\omega$, will lead to a family of equations with our three constants $P_z$, $J_z$, and $C_s$ as parameters.\footnote{In a prior analysis of the rigidly rotating string \cite{Hanna13}, one of these constants ($C_s$) was absorbed into the others, effectively eliminating a scaling degree of freedom.  In the analysis of a flowing skirt \cite{Guven13skirts}, the shapes were presumed to have an additional symmetry, allowing description in terms of a spherical curve rather than a space curve; this yields equations with one less parameter.}   These three constants were found a century ago by Gray \cite{Gray59}, who used them to construct a quadrature akin to \eqref{quadrigid}, but for the radial coordinate $r$ in terms of the coordinate $z$ along the axis of rotation, rather than in terms of the arc length $s$.  

Let us now consider rigid motion of a \emph{shape}, with flow of material along it.  In the case of an inextensible string, this must be a uniform flow along the tangent vectors, as illustrated in Figure \ref{rot} for a rotating shape.  This is a trivial example of a Killing flow, a velocity field on a manifold that preserves its metric.  Such rigid-flowing equilibria occur during the controlled, rapid unspooling of yarn \cite{BatraFraser15}.  Gray discussed properties of such a configuration, but did not construct the solutions \cite{Gray59}.

The symmetry of this system is not in the material coordinate $s$, but in a body-Eulerian coordinate $\eta \equiv s + Tt$, where $T$ is a constant tangential velocity on the string.\footnote{This $T$ is like the $\tau$ of \cite{Guven13skirts}, and should not be confused with the $T$ that briefly appears in \cite{Hanna13} and there represents the total velocity in the direction of the tangents, including any contribution from rigid motion.}  We must consider a body-Eulerian action integral describing a rigidly moving shape into, along, and out of which material is flowing.  We write
\begin{align}
	A &= \int_{t_1}^{t_2} dt \,L_\eta = \int_{t_1}^{t_2} \!\!dt \int_{\eta_1}^{\eta_2} \!\!d\eta \,\mathcal{L}_\eta\left(\bX(\eta,t)\right) \, , \\
	2\mathcal{L}_\eta &= \mu\left(\partial_t\bX + T\partial_\eta\bX\right)\cdot\left(\partial_t\bX + T\partial_\eta\bX\right) - \sigma\left(\partial_\eta\bX\cdot\partial_\eta\bX - 1\right) \, .
\end{align}
The velocity appearing in the kinetic energy term is made up of a piece $\partial_t\bX(\eta,t)$ describing the motion--- in our case, rigid rotation--- of a point on the shape, plus a tangential flow $T\partial_\eta\bX(\eta,t)$ on the shape.  If we were integrating over a material volume, the limits of integration of the $\eta$ integral would be time-dependent, and we would simply be rewriting the previous action.  This is a different action, one that has a symmetry in $\eta$. The situation is reminiscent of many engineering problems involving a finite window of an axially moving system \cite{Renshaw98}, or a garden hose or other conduit \cite{Gay-BalmazPutkaradze14}.

Under the variation $\eta \rightarrow \eta + \delta\eta$, the spatial boundary term is
\begin{align}
	&\quad -\mathcal{L}_\eta + \mu\left(\partial_t\bX + T\partial_\eta\bX\right)\cdot T\partial_\eta\bX - \sigma\partial_\eta\bX\cdot\partial_\eta\bX \, , \\
	&= -\tfrac{1}{2}\mu\partial_t\bX\cdot\partial_t\bX + \tfrac{1}{2}\mu T^2 - \sigma \, , \label{spatialboundary1} \\
	&= -\tfrac{1}{2}\mu\partial_t\bX\cdot\partial_t\bX - \tfrac{1}{2}\mu T^2 - \left(\sigma - \mu T^2 \right)\, . \label{spatialboundary2}
\end{align}
The regrouping of terms between \eqref{spatialboundary1} and \eqref{spatialboundary2} anticipates subsequent expressions.
It will be helpful to rewrite the current in terms of the original material description of $\bX$. Hence,
\begin{align}
	C_\eta - \tfrac{1}{2}\mu T^2 &= \tfrac{1}{2}\mu\partial_t\bX(\eta,t)\cdot\partial_t\bX(\eta,t) + \sigma - \mu T^2 \, , \label{Ceta1} \\
	&= \tfrac{1}{2}\mu\left(\partial_t\bX(s,t) - T\partial_s\bX(s,t)\right)\cdot\left(\partial_t\bX(s,t) - T\partial_s\bX(s,t)\right) + \sigma - \mu T^2 \, . \label{Ceta2}
\end{align} 
The constants $C_\eta$ and $C_s$ coincide when $T=0$.  Note the resemblance of \eqref{Cs} and \eqref{Ceta1} to Bernoulli's equation, applied to ideal fluid flowing along a fixed or rotating streamline.  In fluid or solid, Coriolis terms arising from simultaneous flow and rotation do not contribute to the conserved quantities.\footnote{Healey \cite{Healey96} pointed out the equivalence of one-dimensional solids and fluids in his discussion of axially moving loops of string.}
  Because $\mu T^2$ is a constant, expressions \eqref{Ceta1} and \eqref{Ceta2} are consistent with the (unnumbered) expression for the related constant $C$ derived in a different manner in \cite{Guven13skirts}.
Several results in fluid mechanics, including conservation of potential vorticity and the Kelvin circulation theorem, are derivable either through relabeling symmetry or alternate approaches \cite{Newcomb67, Bretherton70, Salmon88, PadhyeMorrison96-1}.

The other constants are also modified by the flow, and our new choice of action integral.  In terms of the body-Eulerian $\bX(\eta,t)$, 
\begin{align}
	P_z &=  \left[-\mu T\left(\partial_t\bX + T\partial_\eta\bX\right)+ \sigma\partial_\eta\bX\right]\cdot\uvc{z} \, ,  \\
	J_z &= \left[-\mu T\left(\partial_t\bX + T\partial_\eta\bX\right)+ \sigma\partial_\eta\bX\right]\cdot\left(\uvc{z} \times \bX\right) \, , 
\end{align}
and in terms of the material $\bX(s,t)$,
\begin{align}
	P_z &=  \left(-\mu T\partial_t\bX + \sigma\partial_s\bX\right)\cdot\uvc{z} \, , \label{Pzflow} \\
	J_z &= \left(-\mu T\partial_t\bX + \sigma\partial_s\bX\right)\cdot\left(\uvc{z} \times \bX\right) \, . \label{Jzflow}
\end{align}
Equation \eqref{Jzflow} above corresponds with equation (21) in \cite{Guven13skirts}.  Both expressions \eqref{Pzflow} and \eqref{Jzflow} will simplify further after explicit insertion of the material velocity $\partial_t\bX$.

The coordinate system constructed earlier is still useful here.  We now have a shape rigidly rotating with angular frequency $\omega$ about $\uvc{z}$, with an additional uniform flow $T$ along it.  Thus, $\partial_t\bX(s,t) = \omega r \uvc{\theta} + T\partial_s\bX$, and
\begin{align}
	P_z &= \left(\sigma - \mu T^2\right)\partial_s\bX \cdot \uvc{z} \, , \\
	J_z &= \left(\sigma - \mu T^2\right)\partial_s\bX \cdot r\uvc{\theta} - \mu T \omega r^2 \, , \\
	\sigma - \mu T^2 &= C_\eta - \tfrac{1}{2} \mu T^2 - \tfrac{1}{2}\mu\omega^2r^2 \, .
\end{align}
This new expression for $J_z$ already tells us that for nonzero $T$, generic solutions are three-dimensional, rather than coplanar with the axis of rotation.   Note also the expression $\sigma - \mu T^2$, reflecting the Routhian symmetry of inertial frames \cite{Routh55, HealeyPapadopoulos90, Guven13skirts}.  Without the non-inertial frame created by rotation $\omega$, the equilibrium shapes would be unaffected by the flow $T$ along the shape, the centripetal forces arising from this motion balanced merely by a shift in the tension $\sigma$.
Let us rescale using an arbitrary length scale $L$ and the natural parameters of the problem: $s \rightarrow s/L$, $r \rightarrow r/L$, $P_z \rightarrow P_z/\mu\omega^2L^2$, $J_z \rightarrow J_z/\mu\omega^2L^3$, $C_\eta \rightarrow C_\eta/\mu\omega^2L^2$, $T \rightarrow T/\omega L$.  Then, proceeding in a similar manner as before, we obtain
\begin{equation}
	\left(\partial_s r\right)^2 = 1 - \frac{P_z^2 + \left(J_z/r + T r\right)^2}{\left(C_\eta - \tfrac{1}{2} T^2 - \tfrac{1}{2}r^2\right)^2} \, . \label{quadflowing}
\end{equation}
Again, the stress and other coordinates follow directly or by quadratures.
Equation \eqref{quadflowing} represents the full four-parameter family of equilibria of the rotating, flowing string.  This new result extends prior work by several authors \cite{Gray59, Mack58, Caughey69, Fusco84, Hanna13}.\footnote{Note that Mack's \cite{Mack58} solution to this problem cannot be the full solution, because it reduces when $T=0$ to straight lines perpendicular to the rotation axis.  There are many other two-dimensional and three-dimensional equilibria with vanishing $T$ \cite{Gray59, Caughey69, Fusco84, MohazzabiSchmidt99, NordmarkEssen07, Hanna13}.  In writing \cite{Hanna13}, one of the authors misinterpreted Mack's expression in a favorable light--- it does not, in fact, coincide with the general coplanar limit in \cite{Hanna13}, but only with a single special case.  Gray's solution \cite{Gray59} was at that time unknown to the author and, apparently, to the others as well.}
The geometries and stress distributions of the resulting curves are an interesting topic in their own right, and will be the focus of a future paper.

The solutions \eqref{quadflowing} can also be obtained by manipulating variations of multiple auxiliary fields \cite{Guven04, Guven13skirts}, or through extensive calculations involving integrations of projections of equations of motion \cite{Hanna13, Guven13skirts}.
 However, here we have introduced a minimum of variables into the actions, and have not solved or even invoked an equation of motion.  Symmetries and kinematical information are enough to directly reduce the problem to quadratures, using only a few lines of calculation.  Presumably, this approach has general utility, and may be exploited for many other problems involving dynamical equilibria of thin bodies.  
Valuable extensions might include consideration of dissipative dynamics governed by nonconservative action principles \cite{Galley13}, or nonuniform potentials corresponding to the presence of a body force \cite{Kolodner55, AntmanReeken87, LinRavi-Chandar06}.

 \section*{Acknowledgments}

We thank T J Healey and J H Maddocks for helpful comments on a prior draft.  JAH thanks J Guven for introducing him to many of the ideas used in this paper.  JAH acknowledges partial support from U.S. National Science Foundation grant CMMI-1462501.

\newpage

\bibliographystyle{unsrt}

\end{document}